\documentclass[useAMS,usenatbib]{mn2e}
\usepackage{graphicx}
\usepackage{txfonts}
\usepackage{natbib}

\title[X-ray pulsations in IGR J21343+4738]{Discovery of X-ray pulsations in the Be/X-ray binary IGR
J21343+4738}
\author[P. Reig et al.]{P.~Reig$^{1,2}$\thanks{E-mail: pau@physics.uoc.gr},
    A.~Zezas$^{2,1}$	\\
$^{1}$ IESL, Foundation for Research and Technology-Hellas, GR-71110 Heraklion,
Crete, Greece\\
$^{2}$ Institute of Theoretical \& Computational 
Physics, University of Crete, PO Box 2208, GR-710 03, Heraklion, Crete,
Greece }

\newcommand{\igr}  {IGR\,J21343+4738}
\newcommand{\sax}  {SAX\,J2103.5+4545}
\newcommand{\ha}  {H$\alpha$}

\newcommand{\ew}     {EW(H$\alpha$)}

\def\simless{\mathbin{\lower 3pt\hbox
     {$\rlap{\raise 5pt\hbox{$\char'074$}}\mathchar"7218$}}}   
\def\simmore{\mathbin{\lower 3pt\hbox
     {$\rlap{\raise 5pt\hbox{$\char'076$}}\mathchar"7218$}}}   

\def\msun{~{\rm M}_\odot}

\begin{document}

\date{Accepted ??. Received ??; in original form ??}

\pagerange{\pageref{firstpage}--\pageref{lastpage}} \pubyear{2012}

\maketitle

\label{firstpage}

\begin{abstract}

We report on the discovery of X-ray pulsations in the Be/X-ray binary IGR
J21343+4738 during an XMM-Newton observation. We obtained a barycentric
corrected pulse period of $320.35\pm0.06$ seconds. The pulse profile
displays a peak at low energy that flattens at high energy. The pulse
fraction is $45\pm3$\%  and independent of energy within the statistical
uncertainties. The 0.2--12 keV spectrum is well fit by a two component
model consisting of a blackbody with $kT=0.11\pm0.01$ keV and a power law
with photon index $\Gamma= 1.02\pm0.07$. Both components are affected by
photoelectric absorption with a equivalent hydrogen column density
$N_H=(1.08\pm0.15)\times 10^{22}$ cm$^{-2}$ The observed unabsorbed flux is
$1.4\times10^{-11}$ erg cm$^{-2}$ s$^{-1}$ in the 0.2--12 keV energy band. 
Despite the fact that the Be star's  circumstellar disc has almost
vanished, accretion continues to be the main source of high energy
radiation. We argue that the observed X-ray luminosity ($L_{\rm X}\sim
10^{35}$ erg s$^{-1}$) may result from accretion via a low-velocity
equatorial wind from the optical companion.

\end{abstract}


\section{Introduction}

\igr\ was discovered by the {\it INTEGRAL}/IBIS  in December 2002 
\citep{krivonos07,bird07}. The IBIS all-sky hard X-ray survey showed that
\igr\ went through active (December 2002-February 2004) and inactive (Marh
2004-February 2007) periods of X-rays. The
mean X-ray flux during the active period was $(2.3\pm0.4) \times 10^{-11}$
erg cm$^{-2}$ s$^{-1}$ in the 17--60 keV band \citep{bikmaev08}. In
addition to the {\it INTEGRAL} data, there is only one more detection of
\igr\ by an X-ray mission. A short 3.4 ks {\it Chandra} observation of
\igr\ was made in December 2006, that is, when the X-ray emission was below
the threshold of the {\it INTEGRAL} instruments \citep{sazonov08}. Although
the source was clearly detected, the low statistics prevented any detailed
timing or spectral analysis. However, the {\it Chandra} observation allowed
the refinement of its X-ray position and the identification of the optical
counterpart \citep{sazonov08,bikmaev08} 

The association of \igr\ with a Be star identifies this source as a
Be/X-ray binary (BeXB). In this type of X-ray binaries, the normal stellar
component is an Oe or Be star that show spectral lines in emission, while
the X-ray emitting component is a strongly magnetised neutron star
\citep{reig11}. Most BeXB are transient sources with episodes of violent
eruption in X-rays that may reach luminosities of the order of $10^{38}$
erg s$^{-1}$. However, lower luminosity, less variable BeXB also exist.
Transient BeXB usually have short spin  ($P_{\rm spin}\simless 100$ s) and
orbital periods ($P_{\rm orb}\simless 100$ d). Low-luminosity BeXB are
normally persistent sources and tend to have longer spin periods ($P_{\rm
spin}\simmore 100$ s) and wider orbits ($P_{\rm orb}\simmore 100$ d).
Exceptions also occur, such as SAX J2103.5+4545 \citep{hulleman98,baykal00}
and 1A 0118-616 \citep{ives75,staubert11}, which display long pulsation
periods (351 s and 406 s, respectively) and short orbital periods (12.7 d
and 24 d, respectively) and thus do not follow the well-known $P_{\rm
spin}-P_{\rm orb}$ relationship \citep{corbet84}. Although the
identification of \igr\ as a BeXB is firm, neither the orbital or pulse
periods are known.

The nature and optical variability of the massive companion has been amply
investigated by \citet{reig14a}. They found that the optical counterpart to
\igr\ is a $V=14.1$ B1IVe shell star located at a distance of $\sim$8.5
kpc. The long-term optical  variability is characterised by asymmetric
spectral \ha\ line profiles and significant intensity variability. The
changes in the strength of the line are associated with the formation and
dissipation of a circumstellar disc, while those affecting its shape are
believed to be caused by a density perturbation precessing in the disc. The
massive companion in \igr\ is only the second Be shell\footnote{The other
one is 4U\,1258--61\citep{parkes80}. Be shell stars are normal Be stars
seen nearly edge on, that is, with  inclination angles $i \approx
90^{\circ}$ \citep{porter96}.} star in a BeXB.

In this work, we analyse the first {\it XMM-Newton} observations of \igr\
during a low optical state and report the discovery of X-ray pulsations
with a spin period of 320 s.

\section{Observations and data analysis}

\subsection{XMM-Newton observations}

\igr\ was observed by XMM-Newton on 2013 November 24 during revolution
2557. The observation (ObsID 0727961301) started at 18:56 hr UT and lasted
for $\sim$30 ks. The XMM-Newton Observatory \citep{jansen01} includes three
1500 cm$^2$ X-ray telescopes each with an European Photon Imaging Camera
(EPIC) at the focus. Two of the EPIC imaging spectrometers use MOS CCDs
\citep{turner01} and one uses PN CCDs \citep{struder01}. Reflection Grating
Spectrometers \citep[RGS][]{denherder01} are located behind two of the
telescopes while the 30-cm optical monitor (OM) instrument has its own
optical/UV telescope \citep{mason01}. Data were reduced using the
XMM-Newton Science Analysis System (SAS version 13.5).

The EPIC-PN instrument accumulated 0.4--15 keV photons in a {\em full
frame} mode. In this mode, all pixels of all CCDs are read out and thus the
full field of view is covered ($\sim 26 \times 26$ arcmin). The highest
possible time resolution in this mode is 73.4 ms. We used concatenated and
calibrated EPIC event lists available in the PPS Pipeline Products to
generate light curves at different energy. We restricted the useful PN
events to those with a pattern in the range 0 to 4 (single and doubles) and
complying with the more strict selection criterion FLAG=0, which omits
parts of the detector area like border pixels and columns with higher
offset. After filtering, the effective exposure was 27 ks. The
source region is free of pile-up as demonstrated by the fact that the
observed distribution of counts as a function of the PI channel of single
and double events agrees with the expected one. Using the SAS task {\it
epatplot} we find that the 0.5 - 2.0 keV observed-to-model singles and
doubles pattern fractions ratios are consistent with 1.0 within statistical
errors (0.992$\pm$0.027 and 1.011$\pm$0.038 for single and double events,
respectively).

To generate the light curves and spectrum, we extracted events from a
circular region with radius 40 arcsec. To select the background region, we
generated a list of all the detected sources in the PN field of view with
the SAS task {\em edetect\_chain} and chose a source-free region of the
same size $\sim2.5$ arcmin away from the source.

We mainly used data from the EPIC-PN camera because it is the instrument
with the highest effective area.  Nevertheless, because the count rate
below $\sim 1$ keV was very low, we combined data from all three EPIC
cameras for the analysis at low energies, including the search for
pulsations and generation of pulse profile in the energy range 0.3--1 keV,
and the better constraint of the uncertainties in the parameters of the
soft spectral component.

\begin{table}
\begin{center}
\caption{H$\alpha$ equivalent width measurements ($1\sigma$ errors). }
\label{ewha}
\begin{tabular}{cccc}
\hline \hline \noalign{\smallskip}  
Date	&Julian date	&EW(\ha)	 &Telescope	 \\
	&(2,400,000+)	&(\AA)		 &		 \\

\hline \noalign{\smallskip}
04-01-2013 &56297.57	&$-2.1\pm0.2$	&FLW \\
18-10-2013 &56584.31	&$+2.1\pm0.2$	&SKO \\
03-11-2013 &56599.63	&$+2.1\pm0.2$	&FLW \\
08-12-2013 &56634.61	&$+2.6\pm0.2$	&FLW \\
\hline \noalign{\smallskip}
\end{tabular}
\end{center}
\end{table}

\begin{figure}
\resizebox{\hsize}{!}{\includegraphics{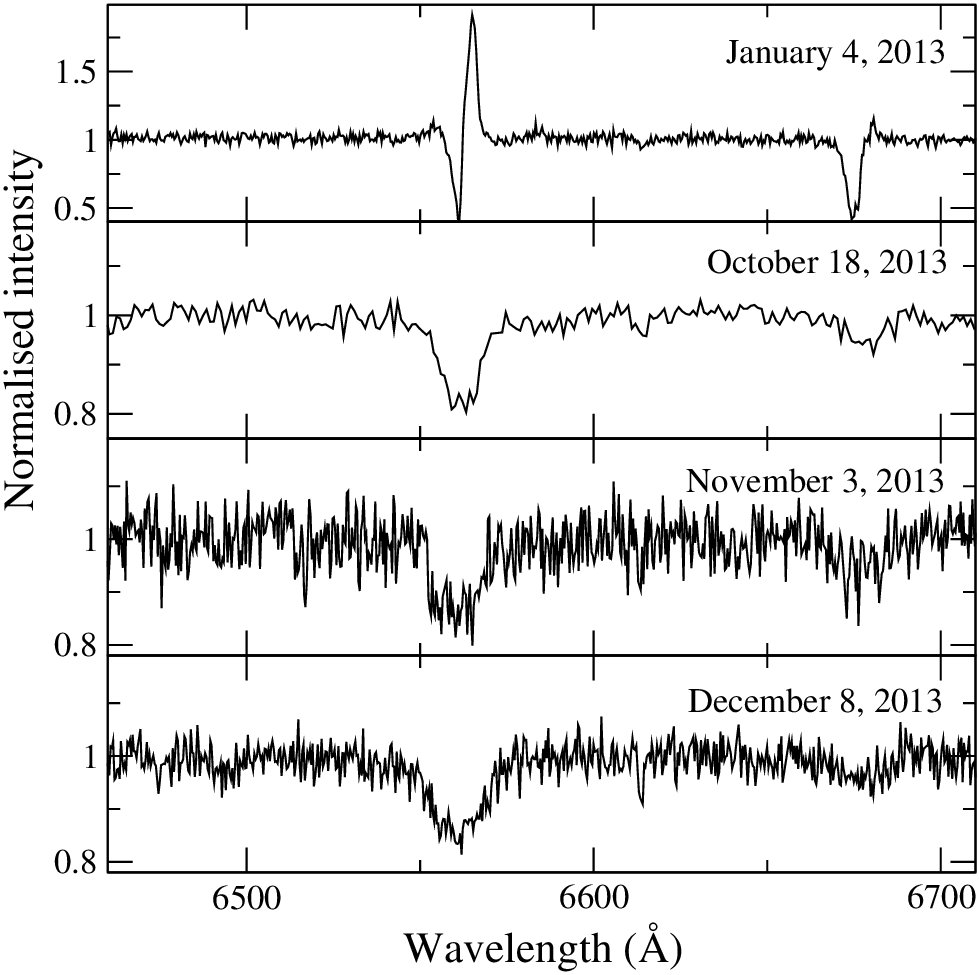} } 
\caption[]{\ha\ line profile. The {\it XMM-Newton} observations took place
in November 24, 2013.}
\label{opt-spec}
\end{figure}

\subsection{Optical spectra}

A long-term optical study of the donor star in \igr\ has been presented in
\citet{reig14a}.  Here we present new optical spectroscopic observations
that demonstrate the absence of optical emission at the time of the X-ray
observations. The dominant formation process of the hydrogen emission lines
is recombination of the stellar flux by the disc. Thus, in the absence of
the disc, no emission should be observed and the line should display the
normal photospheric absorption profile. By convention, the equivalent width
of an emission line is negative, while a positive value indicates an
absorption profile.   Although there is no  optical observation at the
exact time of the {\it XMM-Newton} observation, an absorption profile is
observed 15--20 days before and after. This time is not enough for a disc
to form.  The formation and dissipation of a circumstellar disc in Be stars
happen on timescales of months to years \citep{jones08}.

The optical spectra were obtained from the Skinakas observatory (SKO) in
Crete (Greece) on the night October 18, 2013 and from the Fred Lawrence
Whipple observatory (FLWO) at Mt. Hopkins (Arizona) on the night November
3, 2013 and December 8, 2013. We have also included our last observation in
which the star was seen with  the \ha\ line in emission. The 1.3\,m
telescope of the Skinakas Observatory was equipped with a 2000$\times$800
(15 $\mu$m) pixel ISA SITe CCD and a 1302 l~mm$^{-1}$ grating, giving a
nominal dispersion of $\sim$1 \AA/pixel.  The FLW observations  were made
with the 1.5-m  telescope, 1200 l~mm$^{-1}$ grating and the FAST-II
spectrograph \citep{fabricant98}. 

Table~\ref{ewha} gives the log of the spectroscopic observations and the
value of the H$\alpha$ equivalent width. Fig.~\ref{opt-spec} shows the line
profiles.


\begin{figure}
\resizebox{\hsize}{!}{\includegraphics{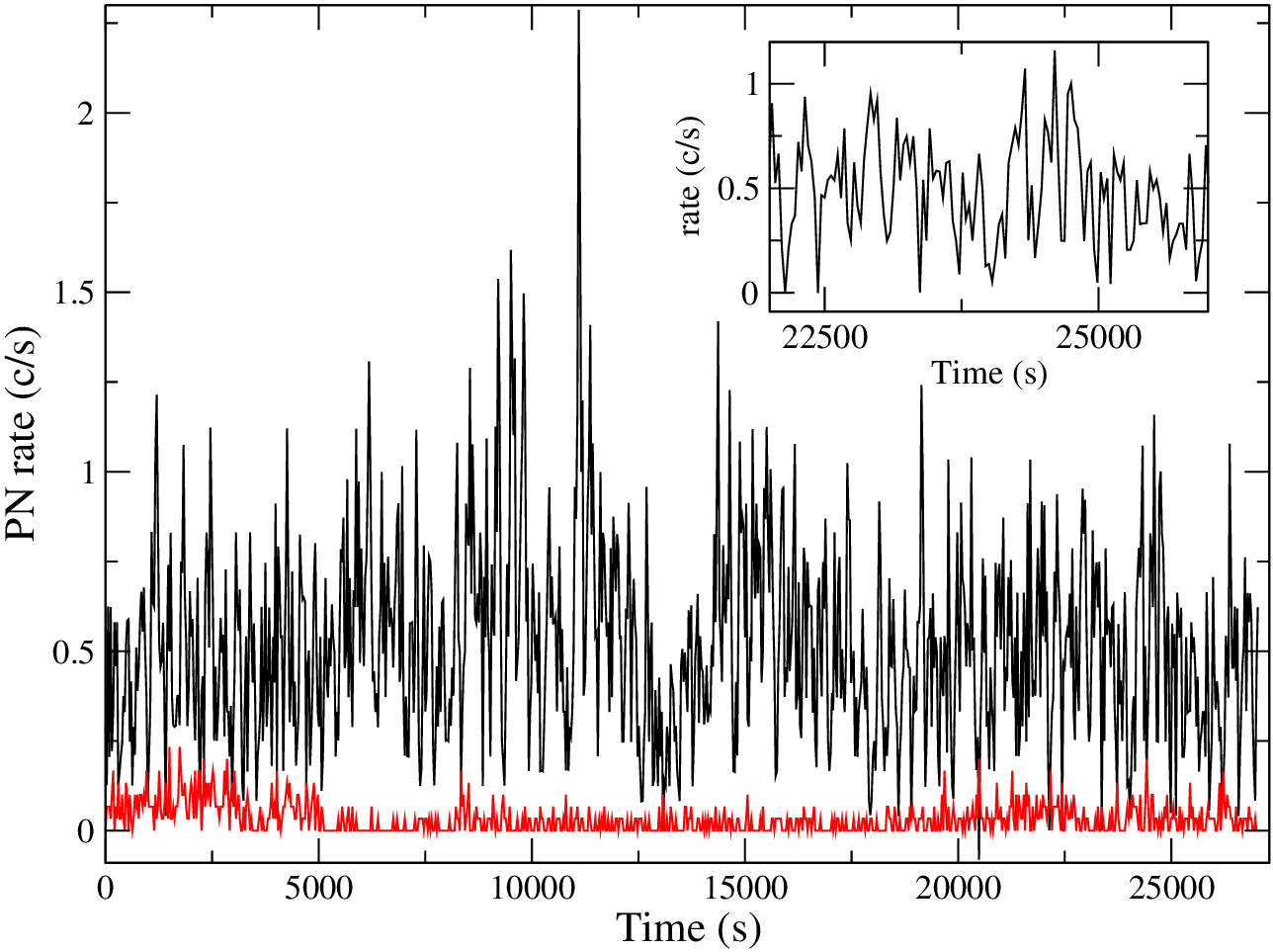} } 
\caption[]{Background subtracted EPIC-PN light curve (black) and background
light curve (red) in the energy range 0.2--12 keV with time bin of 30 s. 
The inset shows the pulsed nature of the X-ray emission.}
\label{pn-lc}
\end{figure}

\section{Results}

\subsection{Timing analysis and detection of X-ray pulsations}

The PN light curve with 30 s time resolution in the energy range 0.2-12 keV
is shown in Fig.~\ref{pn-lc} for the entire duration of the observation and
for a shorter interval (see inset), where the presence of pulsed emission
can be seen. In addition to the pulsations, the X-ray light curve shows
moderate variability on time scales of a few ks. The root-mean-square
measured in the 30 s binned light curve is $44\pm2$\%. The average
source count rate after background subtraction is
$0.517\pm0.005$ counts s$^{-1}$. The average background rate for the entire
27 ks observation is $0.033\pm0.001$ counts s$^{-1}$, but it is about 35\%
higher at the beginning of the observation owing to enhanced flaring
particle background. If the first 5000 s of data are removed, then the
average background count rate reduces to $0.024\pm0.001$ counts s$^{-1}$.

\begin{figure}
\resizebox{\hsize}{!}{\includegraphics{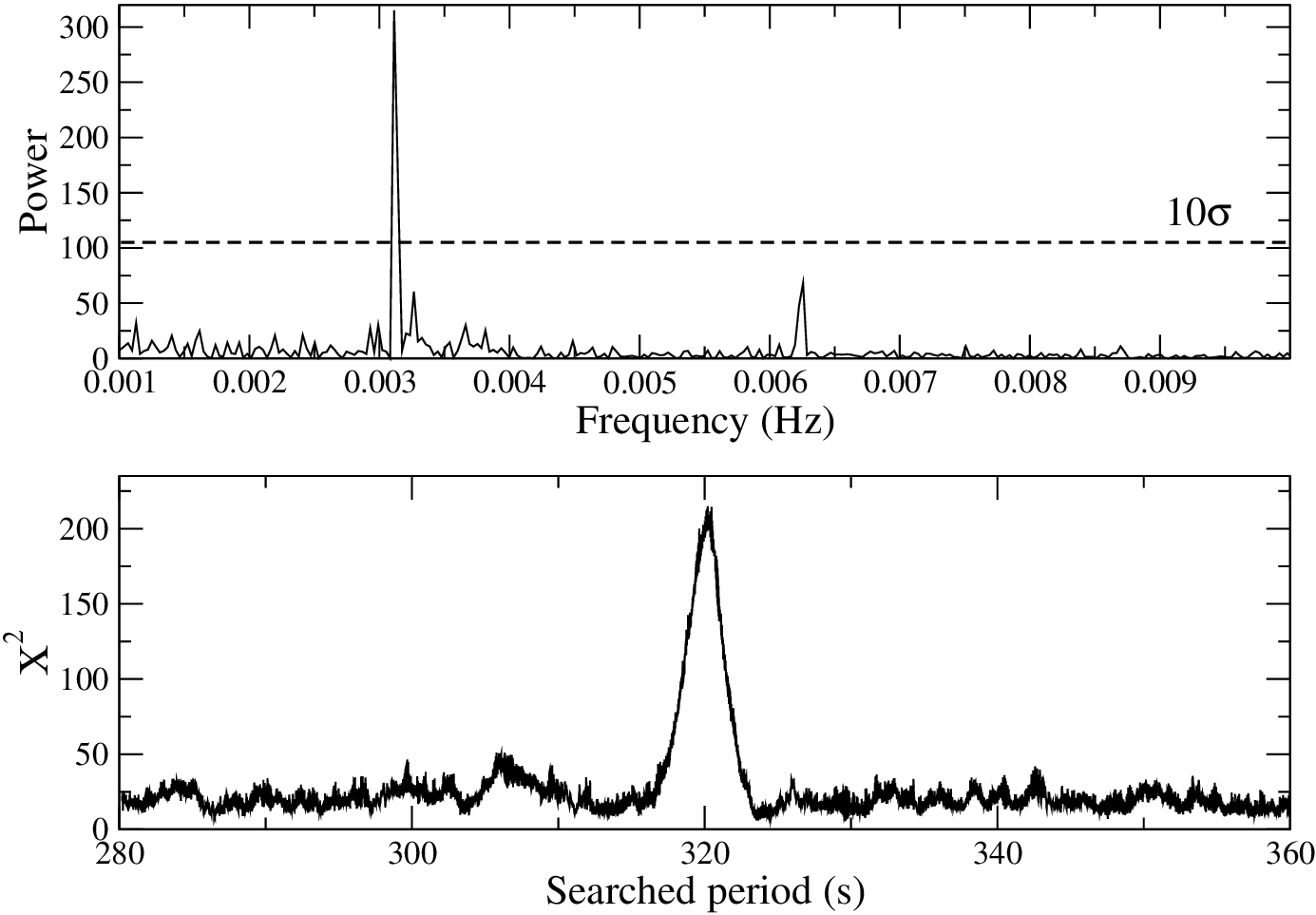} } 
\caption[]{{\em Upper panel}: the EPIC-PN power spectrum and $\sim10\sigma$
significance level. {\em Lower panel}: $\chi^2$ maximization after folding
the data over a range of periods (epoch folding).}
\label{spin}
\end{figure}

The timestamps of each event were corrected to the Earth's barycenter prior
to the extraction of the light curves from the event file. To search for
periodic signal, we generated a  0.2--12 keV light curve with a time
resolution of 1 s. Then we run a fast Fourier transform (FFT) algorithm to
produce a single Leahy-normalized \citep{leahy83} power spectrum covering
the frequency interval $3.05 \times 10^{-5}-0.5$ Hz in 16384 bins
(Fig.~\ref{spin}). The maximum power is distributed over two bins at a
frequency 0.003128052 Hz, which corresponds to a periodic signal with a
period of $\sim 320$ s. Another strong peak is seen at twice this frequency
and correspond to the first harmonic of the signal. The dashed line in the
upper panel of Fig.~\ref{spin} represents the $\sim10\sigma$ detection
level.

In addition, we also searched for periodic signal using an epoch-folding
technique where the data is folded over a period range
\citep{leahy87,larsson96}. For each trial period the $\chi^2$ statistic in
calculated. If the data contain a periodic signal, then a peak stands out
in the $\chi^2-P_{\rm trial}$ plot. We used the task {\it efsearch} of the
XRONOS package and found a best-fit period at $320.4\pm0.1$~s
(Fig~\ref{spin}, lower panel). 

To improve the estimation of the pulse period, we next applied a phase
fitting technique. We divided the light curve into 8 segments, each of one
with a length equal to 10 pulse periods and calculated a folded pulse
profile for each segment with a common epoch and period. These profiles
were cross-correlated with a template obtained by folding the entire light
curve onto the trial period. The resulting phase delays were fitted with a
linear function, whose slope provides the correction needed to be apply to
the trial period. We then adjusted the period and repeated the procedure
until the phase delay exhibited no net trend with time throughout the
observation.  The best-fit period was $320.35\pm0.06$~s, where the error
was estimated from the uncertainty on the first-order term of a linear fit
to the phase delays.

\begin{figure}
\resizebox{\hsize}{!}{\includegraphics{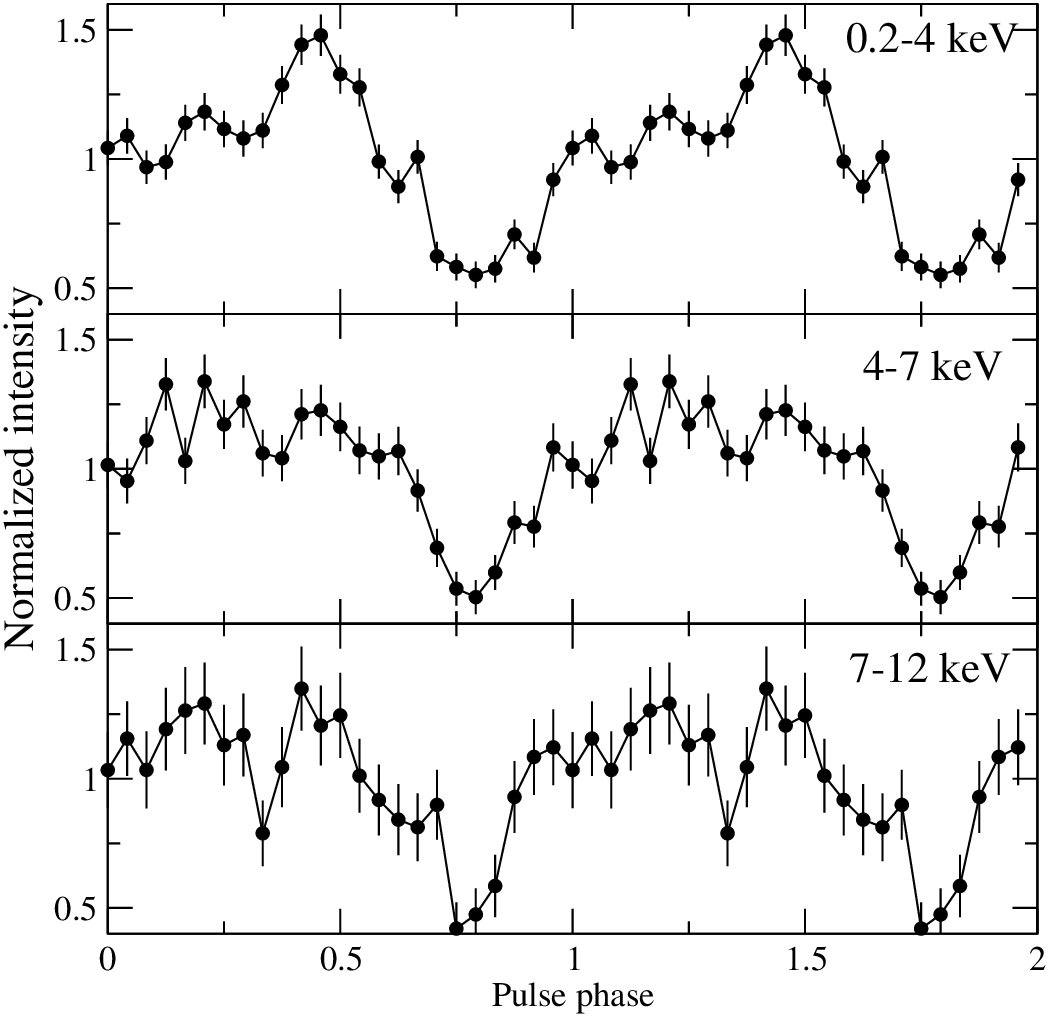} } 
\caption[]{Normalized pulse profiles at different energies.}
\label{profile}
\end{figure}

The pulse profiles obtained by folding light curves at different energy
with the best-fit period are shown in Fig.~\ref{profile}.  The profile
exhibits a peak at the pulse maximum at low energies that disappears at
high energies. At $\simmore 4$ keV the pulse maximum is flat. The pulse
fraction, defined as $PF = (I_{\rm max}-I_{\rm min})/(I_{\rm max}+I_{\rm
min})$, where $I_{\rm min}$ and $I_{\rm max}$ are background-corrected
count rates at the pulse profile minimum and maximum, is $75\pm15$\% below
1 keV but remains constant within the errors above that energy at a level 
of $45\pm3$\%.

\begin{table}
\begin{center}
\caption{PN best-fit spectral parameters. A single power law (PL) and a power law 
plus a blackbody (BB) were considered. Errors correspond to 90\% confidence. 
Quoted luminosity for an assumed distance of 8.5 kpc.}
\label{specfit}
\begin{tabular}{lll}
\hline \hline \noalign{\smallskip}
Parameter			&Best-fit value		&Best-fit value\\
				&(PL+BB)		&(PL) \\
\hline \noalign{\smallskip}
$N_{\rm H}$			&1.08$\pm$0.15		&0.63$\pm$0.06\\
$kT_{\rm BB}$ (keV)		&0.11$\pm$0.01		&--\\
Photon index			&1.02$\pm$0.07		&0.85$\pm$0.05\\
$F^{\rm unabs}_{\rm 0.2-12 \, keV}$ (erg cm$^{-2}$ s$^{-1}$)&$1.4\times 10^{-11}$	&$5.8\times 10^{-12}$\\	
$L^{\rm unabs}_{\rm 0.2-12 \, keV}$ (erg s$^{-1}$)	&$1.2\times 10^{35}$	&$5.0 \times 10^{34}$\\	
$L^{\rm abs}_{\rm 0.2-12 \, keV}$ (erg s$^{-1}$)	&$4.4\times 10^{34}$	&$4.5 \times 10^{34}$\\	
$\chi^2_r$/dof			&1.13/111		&1.39/113\\
\hline \hline \noalign{\smallskip}
\end{tabular}
\end{center}
\end{table}

\subsection{Spectral analysis}

We extracted a PN energy spectrum using the same source and background
regions as for the timing analysis and filtering criteria described. For
the spectral analysis we further cleaned the data by accepting only the
good times when sky background was low.  In particular, the first
$\sim5000$ seconds of the observations, when enhanced background is
observed, were discarded.  We rebinned the energy spectra by requiring at
least 50 counts for each energy bin. 

\begin{figure}
\resizebox{\hsize}{!}{\includegraphics{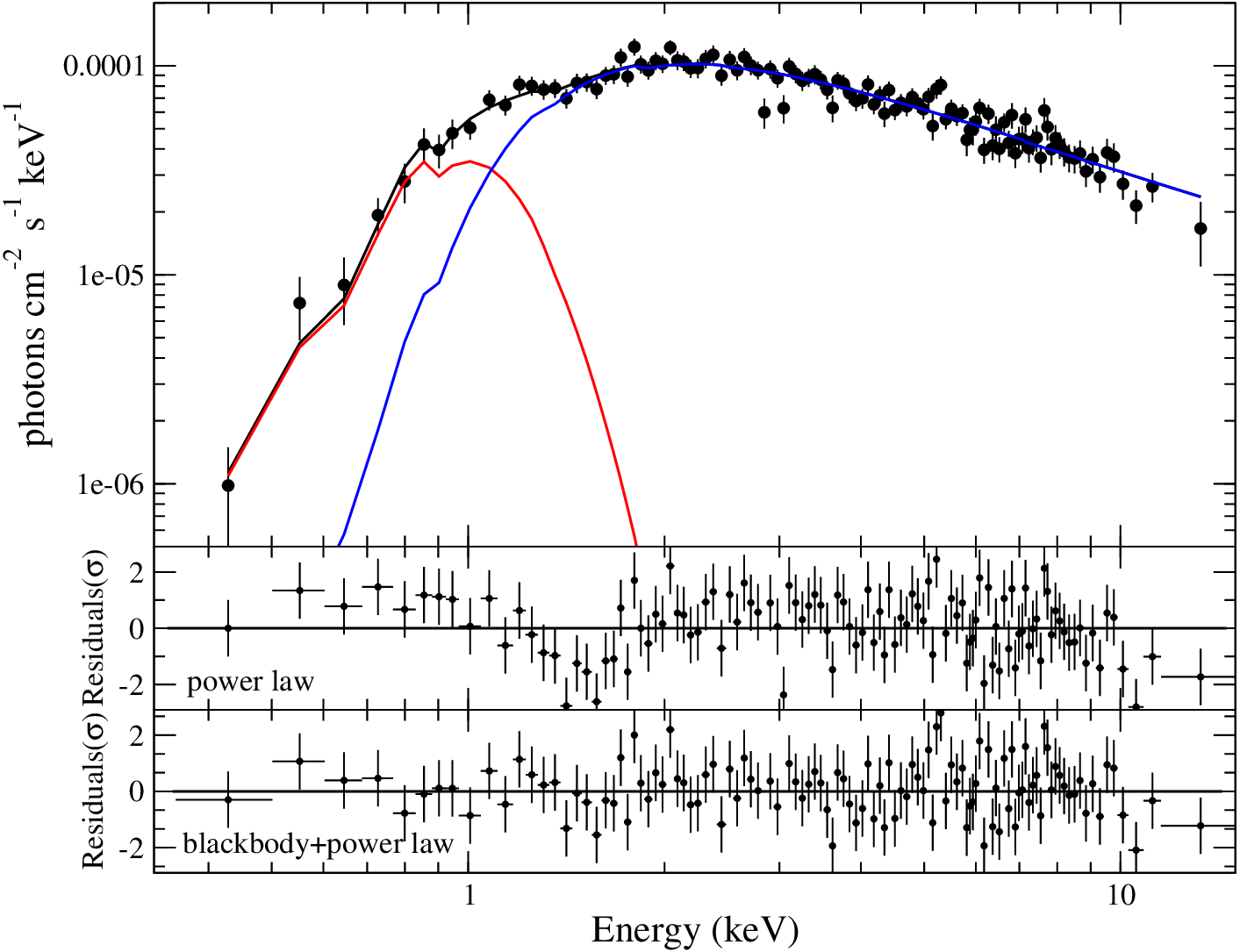} } 
\caption[]{PN spectrum of \igr\ (circles) and best-fit model
(black line), which consists of a blackbody (red line) and a power law 
(blue line) components. The middle panel represents the residuals in terms
of sigmas without the blackbody component. The bottom panel gives the
residuals when the blackbody component is added to the fit.}
\label{pn-spec}
\end{figure}

A single component model composed of an absorbed power-law does not
provide a satisfactory fit ($\chi^2=157$ for 113 degrees of freedom),
leaving a deficit of flux at around 1.6 keV and weak soft excess below 1
keV (see Fig.~\ref{pn-spec}).  The addition of a blackbody component
reduces significantly the amplitude of these two features and improves
significantly the fit  ($\chi^2=125$ for 111 degrees of freedom).  
Table~\ref{specfit} summarises the results of the spectral fit. The column
density was obtained assuming abundances given in \citet{anders89} and
crosssections from \citet{balucinska-church92}, with improvements from
\citet{yan98}.

However, the normalization of the blackbody component is poorly constrained
(relative error of $\sim$50\%). Leaving only the normalization of the
blackbody and power law components free and fixing the rest of the
parameters to their best-fit values, the uncertainty in the blackbody
normalization goes down to $\sim$15\%. In an attempt to better constrain
the parameters of the blackbody component, we combined the spectra of all
the EPIC cameras into a single spectrum. The best-fit parameters agree well
with those obtained from the PN spectrum alone within errors, namely
$N_H=(1.2\pm0.1)\times 10^{22}$ cm$^{-2}$, $kT=0.10\pm0.01$ keV, and
$\Gamma=1.10\pm0.05$. In this case, the uncertainty in the blackbody
normalization reduces to 9\% and the reduced $\chi^2_r=1.06$, for 484 degrees
of freedom.

In a paper devoted to cross-calibration of X-ray instruments,
\citet[][see Fig. 15]{tsujimoto11} found excess of emission below 1.5 keV
in the {\it XMM-Newton}/PN spectrum of the calibrating source G21.5-0.9.
The origin of this feature is unknown. Although there are important
differences\footnote{The calibrating source G21.5-0.9 is an extended X-ray
source that fell in more than one PN chip. The background and chip gap
correction issues do not affect \igr. Likewise, the instrumental effect
appears to peak at above 1 keV, whereas the soft excess in \igr\ becomes
apparent below 1 keV.} between G21.5-0.9 and \igr, the significance of the
blackbody component should be treated with caution. 

We found no evidence for an iron line emission line. The upper limit on the
equivalent width of a 6.4 keV emission line is 28 eV.

The absorption significantly affects the shape of the spectrum below $\sim$
2 keV. The absorbed and unabsorbed X-ray fluxes in the 0.2--12 keV energy
range are $5.1\times 10^{-12}$ erg cm$^{-2}$ s$^{-1}$ and $1.4\times
10^{-11}$ erg cm$^{-2}$ s$^{-1}$, respectively, whereas these fluxes for
the 2--10 keV energy range are  $3.8\times 10^{-12}$ erg cm$^{-2}$ and
$4.1\times 10^{-12}$ erg cm$^{-2}$, repectively. Assuming a distance of 8.5
kpc \citep{reig14a}, these values correspond to an absorbed and unabsorbed
luminosities of $4.4\times 10^{34}$ erg s$^{-1}$ and $1.2\times 10^{35}
$erg s$^{-1}$, respectively, in the 0.2--12 keV band. In the case of the
single power-law model, the  column density is lower and the  unabsorbed
luminosity is  $5.0\times 10^{34}$ erg s$^{-1}$. Extra uncertainty in the
luminosity comes from the  uncertainty in the distance estimation. Taking
all this into account, we conclude that the X-ray luminosity ranges between
$(0.5-1.5)\times 10^{35}$ erg s$^{-1}$. Fig.~\ref{pn-spec} shows the PN
spectrum and best fit, together with the residuals with (bottom panel) and
without (middle panel) the addition of the blackbody component.

\section{Discussion}

We have performed an X-ray timing and spectral analysis of \igr\ and  shown
for the first time that the X-ray emission is pulsed with a pulse period of
320 s and a pulse fraction of $\sim45$\%. Although the X-ray information of
\igr\ is scarce, the discovery of coherent X-ray pulsations is a key
element to understand the nature of the system. Most X-ray pulsars are in
high-mass X-ray binaries. The association of the X-ray source \igr\ with a
Be star, initially suspected by \citet{bikmaev08} has been firmly confirmed
by \citet{reig14a}, who suggested a $V=14.1$ B1IVe shell star located at a
distance of $\sim$8.5 kpc as the most likely optical counterpart. Virtually
all BeXBs harbour accretion-powered pulsars. Thus  the discovery of X-ray
pulsations reinforces the membership of \igr\ to the group of BeXB. 
However, the  orbital period of \igr\ remains unknown.  According to the
$P_{\rm spin}-P_{\rm orb}$ relationship \citep{corbet84}, a $P_{\rm
spin}=320$ s implies that an orbital period in the range 200--300 days
should be expected. Such a long orbital period and most likely persistent 
X-ray emission at a level of $\simless 10^{35}$ erg s$^{-1}$ require
the orbit to be nearly circular and accretion to occur via wind material
from the massive companion in a similar fashion to the sub-class of BeXB
suspected to be born with a smaller kick velocity ($\simless 50$ km
s$^{-1}$) after the supernova explosion that gave birth to the neutron
star. The characteristic of this sub-class of BeXBs has been studied by
\citet{Pfahl02}. A natural consequence of a wide and circular orbit is the
formation of extended and large circumstellar discs because the neutron
star would not truncate the disc until it grows to a certain size. Thus,
large \ha\ equivalent widths should be expected in this type of systems.
Indeed, members of this subclass of BeXB, such as X-Per, GS\, 0834--43,
KS\,1947+300, and 1RXS J225352.8+624354 have \ha\ equivalent widths, \ew\
$\simmore -15$ \AA\ \citep{clark01,israel00,negueruela03,esposito13}. In
contrast, the maximum \ew\ measured in \igr, in observations spaced by
several years, is $-8$ \AA\ \citep{reig14a}. This weak \ha\ emission points
to a small and compact circumstellar disc, which could be the result of
truncation by the neutron star. Truncation would require a narrower orbit
and a shorter orbital period than those observed in the near-circular orbit
systems.

Although \igr\ is a variable source, it has not so far shown large type II
outbursts as in BeXBs transients with much faster rotating neutron star
such as 4U\,0115+65 \citep{li12,muller13} or V\,0332+54
\citep{tsygankov10,nakajima10}. In fact, the {\it Chandra} detection
\citep{sazonov08} when the source was thought to be in an off-state (below
the threshold of the {\it INTEGRAL} instrument) suggests the possibility
that \igr\ is a persistent source.  \igr\ could be similar to the BeXB
\sax, which also displays weak \ha\ emission --- the maximum \ew\ ever
reported is only $\sim -5$ \AA\ \citep{reig10a}.  \sax\ has a spin period
of 355 seconds \citep{hulleman98} and an orbital period of 12.7 days
\citep{baykal00} and shows extended bright and faint X-ray states that last
for a few hundred days \citep{reig10a}. During the bright states, the
neutron star spins up \citep{baykal07,camero07}, shows moderate type I
outbursts modulated  by the orbital period \citep{baykal00,sidoli05}, and
displays enhanced optical emission, characterised by \ha\ line emission
\citep{kiziloglu09,reig10a}. During the faint state,  the modulation of the
X-ray intensity with orbital phase disappears \citep{baykal02,blay04}, the
spin frequency of the neutron star remains fairly constant \citep{baykal07}
or slightly decreases, i.e., it spins down \citep{ducci08} and the B-type
companion shows \ha\ in absorption \citep{kiziloglu09,reig10a}. The average
X-ray luminosity in the low X-ray state of \sax\ is $L_X\simless 10^{35}$
erg s$^{-1}$ in the 3--30 keV range. In short, the X-ray (long-period
pulsations, bright and faint states, very weak iron emission line) and
optical (small \ew, fast spectral changes) variability of \igr\ resembles
that of the persistent source \sax.

Our {\it XMM-Newton} observations of \igr\ took place during a low optical
state, when the \ha\ line was in absorption, implying the disappearance of
the Be star's circumstellar disc (Fig.~\ref{opt-spec}).   The source
has been in this state over more than six months \citep{reig14a}. All
spectra taken from June to December 2013 show an absorption profile.
Without the disc,  accretion  should cease because the material in the disc
constitutes the fuel that powers the X ray emission. Still, pulsations are
clearly detected, indicating that the accretion mechanism remains active.
The measured  X-ray luminosity is comparable to other persistent BeXB
and well above the minimum luminosity at which the propeller effect sets
in. For \igr, using equation (1) in \citet{campana02} and the canonical
mass and radius of a neutron star and a typical magnetic field of $2 \times
10^{12}$ G, we obtain $L_{\rm min}=2.2 \times 10^{32}$ erg s$^{-1}$.  In
the absence of the disc, X-rays could result from accretion from a stellar
wind. Even though the wind from a main-sequence star is not as powerful as
in a supergiant star, a narrow and eccentric orbit could in principle
explain the observed luminosity, as it has been proposed for \sax\
\citep{reig05a}.

Assuming that all the gravitational energy is converted into X-rays, the
X-ray luminosity powered by the stellar wind is given by \citep[see
e.g.][]{waters89a} 

\begin{eqnarray}
\nonumber
 L_{\rm wind} \approx& 4.8 \times 10^{37} 
\left(\frac{M_{\rm X}}{1.4 {\rm \msun}}\right)^3 
\left(\frac{R_{\rm X}}{10^6 \,{\rm cm}}\right)^{-1}\!
\left(\frac{M_*}{1 {\rm \msun}}\right)^{-2/3} \!
\left(\frac{P_{\rm orb}}{1 \, {\rm day}}\right)^{-4/3} \! \\ 
&\left(\frac{\dot{M}_*}{10^{-6} {\rm \msun yr^{-1}}}\right) \!
\left(\frac{v_{\rm w}}{10^8 \,{\rm cm \, s^{-1}}}\right)^{-4} \!
 {\rm erg \, s^{-1}} 
\end{eqnarray}

\noindent where $M_{\rm X}$ and $R_{\rm X}$ are the mass and radius of the
neutron star, $M_*$ and $\dot{M}_*$ the mass and mass loss rate of the
optical companion, $P_{\rm orb}$ the orbital period of the binary, and
$v_{\rm w}$ the stellar wind velocity.   The wind parameters of
main-sequence Be stars are rather uncertain. Typical parameters for an
early-type Be star are $M_*\sim12 \msun$, $\dot{M}_* \sim (2-5) \times
10^{-8}$ $\msun$ yr$^{-1}$, $v_{\rm w}\sim (0.8-1) \times 10^8$ cm s$^{-1}$
\citep{prinja89}. However, these wind velocities were derived from
ultraviolet observations, hence represent polar wind velocities. Infrared
observations give a different picture and support the existence of a
lower-velocity higher density equatorial wind, where the wind velocities
are of the order of 150--600 km s$^{-1}$ and mass-loss rates an order of
magnitude higher \citep{waters88}. 

Assuming a mass-loss rate in the range $\dot{M}_* \sim (1-9) \times
10^{-8}$ $\msun$ yr$^{-1}$, a velocity typical of the equatorial wind,
$\sim 300-500$ km s$^{-1}$ would reproduce the  observed unabsorbed X-ray
luminosity of $10^{35}$ erg s$^{-1}$\ in \igr. However, this would imply
that the disc has not completely vanished, in contradiction with the
profile of the \ha\ line. This apparent contradiction can be solved taking
into account the formation loci of the different observables in Be stars.
Models of gaseous discs in Be stars  show that disc emission only fills in
the photospheric absorption profile when the disc size is about 5 stellar
radii \citep{carciofi11}. Thus even if an absorption profile is seen, a
small disc might still be present.  Two interesting facts support the
presence of a small disc. First, the typical \ha\ equivalent width for a
normal (non-emitting) early-type B stars is about +3.5 \AA\
\citep{jaschek87}, whereas the measured equivalent width in \igr\ is well
below +3 \AA, suggesting a small amount of fill-in emission. Second, the
\ha\ line shape of the October and November 2013 spectra
(Fig.~\ref{opt-spec}) is reminiscent of the so-called central
(quasi)-emission peak  profile, which appear when the innermost regions of
the disc are being supplied with matter \citep{rivinius99}. Such a disc
must be small and tenuous because no iron line at 6.4 keV is detected. This
line results from reprocessing of the X-ray continuum in relatively cool
matter and it a very common if not ubiquitous feature in BeXB
\citep{reig13}. Its detection  provides strong evidence for the presence of
material in the vicinity of the X-ray source.  

The 0.2--12 keV  spectrum of \igr\ is well represented by a power law model
and a blackbody component affected by photoelectric absorption.  The
blackbody component accounts for the excess of emission $\simless 1$ keV. 
Soft excess is a common feature observed in many, probably all, 
accreting X-ray pulsars
\citep{hickox04,mukherjee05,lapalombara06,lapalombara07,lapalombara09}.
\citet{hickox04} showed that when the soft excess is modelled with a
blackbody component, the blackbody spectral parameters (temperature and
emission radius) strongly depend on the source luminosity, which in turn,
depends on the accretion mechanism.  In high-luminosity pulsars, such as
SMC X-1, Cen X-3 and LMC X-4, accretion occurs via Roche lobe overflow and
an accretion disc is formed. In these systems $kT\approx  0.1$ keV and
$R_{\rm bb}\approx 1000$ km, typically, and the soft excess is believed to
be due to reprocessing of hard X-rays in the inner parts of an accretion
disc.  Low-luminosity pulsars have $kT \approx 1-1.5$ keV and $R_{\rm
bb}\approx 0.1$ km. In these systems the soft excess would be originated at
the base of the accretion column, close to the neutron star surface. The
wind-fed supergiant systems have $kT\approx0.2$ keV and $R_{\rm bb} \approx
60-100$ km. In these systems, the thermal component responsible for the
soft excess is likely to be a cloud of diffuse plasma around the neutron
star. In \igr, the best-fit gives a blackbody temperature and radius  of
the emitting region\footnote{The normalization of the blackbody component
is $N=R_{\rm bb}^2/D_{10}^2$, where $R_{\rm bb}$ is the source  radius in
km, and $D_{10}$ is the distance to the source in units of 10 kpc.} are 
$kT=0.11\pm0.01$ K and $R_{\rm bb}=80\pm15$ km, for a distance of 8.5 kpc.
The emitting region responsible for the soft excess in \igr\ is too large
to come from the polar cap of the neutron star and too small to come from
the inner parts of an accretion disc. 
The radius of the accreting polar cap is estimated as $R_{\rm cap}\approx
R_{\rm X}(R_{\rm X}/R_{\rm m})^{1/2}$, where $R_{\rm m}$ is the radius of
the magnetosphere, which for the case of \igr, i.e. assuming $L_{\rm X}=1
\times 10^{35}$ erg s$^{-1}$ and a typical magnetic moment of $10^{30}$ G
cm$^3$, is $R_{\rm m}=1.4\times 10^9$ cm. Thus $R_{\rm cap}\sim 0.27$ km.
On the other hand, the inner parts of an accretion disc would be outside
the magnetosphere ($>14000$ km). The fact that pulsations are still
detected below $\sim$1 keV indicates that the blackbody emitting region is
phase locked with the neutron star.

An alternative source of matter during the absence of the circumstellar
wind might be a residual accretion disc around the neutron star  as has
been suggested for the BeXB 1A\,0535+262 \citet{doroshenko14}, based on the
similarity of the spectral and timing properties of the quiescent  and
outburst X-ray emission. Because there is evidence that an accretion disc
is present in 1A\,0535+262 during outbursts, the similarities suggest a
common origin of the quiescent and bright X-ray emission. Given that no
X-ray outburst has been seen in \igr, no such comparison is possible. 
Note however, that the absence of the iron emission line argues against
large amount of matter (e.g. an accretion disc) in the vicinity of the
system.

When only high-energy data from {\it INTEGRAL} was available, \igr\
was seen to go through on- and off X-ray states as many other transient
Be/X-ray pulsars.  The {\it INTEGRAL} luminosity during the active states
was comparable to the one detected by {\it XMM-Newton} in this work, namely
in the order of $10^{35}$ arg s$^{-1}$.  The {\it Chandra} detection when
the source was in an off state and the {\it XMM-Newton}
observation during a dis-loss episode suggest that \igr\ is a persistent
source. Moreover, the lack of large amplitude  X-ray intensity variability
(no giant X-ray outbursts typical of transient systems have been
seen in \igr) and the lack of a strong iron line are also characteristics of
persistent BeXB \citep{reig99}. 
Nevertheless, the number of X-ray
observations of \igr\ is still very limited and larger amplitude increases
of X-ray intensity in the form of outbursts cannot be ruled out. An example
of a persistent BeXB which suddenly exhibited enhanced X-ray activity is
RX\,J0440.9+4431 \citep{ferrigno13}.

\section{Conclusion}

We have performed the first detailed X-ray timing and spectral analysis of
the Be/X-ray binary \igr\ and found that its X-ray emission pulses with a
pulse period of 320 s.  The X-ray observation took place at a time when the
B star companion had (almost) lost the circumstellar disc, which it is
assumed to be the prime source of matter that powers the X-ray through
accretion. Despite this low-optical state, the  X-ray luminosity is
comparable to that seen in other persistent Be/X-ray binaries with discs
and significantly larger than the minimum luminosity at which accretion is
expected to halt due to the centrifugal inhibition caused by the propeller
effect. We argue that a equatorially enhanced stellar wind could
explain the observed luminosity. The X-ray spectrum shows a soft excess,
whose origin is not clear. If it is modeled with a blackbody, then the size
of the emitting region is not compatible neither with the polar cap nor the
inner parts of an accretion disc.

\section*{Acknowledgments}

This work has made use of NASA's Astrophysics Data System
Bibliographic Services and of the SIMBAD database, operated at the CDS,
Strasbourg, France.
 
\bibliographystyle{./mn2e}
\bibliography{ref-x}

\end{document}